\documentclass[aps,prb,twocolumn,superscriptaddress,10pt]{revtex4-2}

\usepackage{mathtools}
\usepackage{braket}
\usepackage{tikz}
\usetikzlibrary{arrows,arrows.meta,calc}
\usepackage[nolist,nohyperlinks]{acronym}
\usepackage{subcaption}
\usepackage{siunitx}
\usepackage{lineno}
\usepackage[draft]{changes}
\usepackage{bbold}
\usepackage{hyperref}
\hypersetup{
    colorlinks=true,
    linkcolor=blue,
    filecolor=magenta,      
    urlcolor=cyan,
    pdftitle={Overleaf Example},
    pdfpagemode=FullScreen,
    }

\usepackage{xcolor}
\definecolor{halfgray}{gray}{0.55}
\definecolor{ipython_frame}{RGB}{207, 207, 207}
\definecolor{ipython_bg}{RGB}{247, 247, 247}
\definecolor{ipython_red}{RGB}{186, 33, 33}
\definecolor{ipython_green}{RGB}{0, 128, 0}
\definecolor{ipython_cyan}{RGB}{64, 128, 128}
\definecolor{ipython_purple}{RGB}{170, 34, 255}

\usepackage{listings}
\lstdefinelanguage{iPython}{
    morekeywords={access,and,break,class,continue,def,del,elif,else,except,exec,finally,for,from,global,if,import,in,is,lambda,not,or,pass,print,raise,return,try,while},%
    morekeywords=[2]{abs,all,any,basestring,bin,bool,bytearray,callable,chr,classmethod,cmp,compile,complex,delattr,dict,dir,divmod,enumerate,eval,execfile,file,filter,float,format,frozenset,getattr,globals,hasattr,hash,help,hex,id,input,int,isinstance,issubclass,iter,len,list,locals,long,map,max,memoryview,min,next,object,oct,open,ord,pow,property,range,raw_input,reduce,reload,repr,reversed,round,set,setattr,slice,sorted,staticmethod,str,sum,super,tuple,type,unichr,unicode,vars,xrange,zip,apply,buffer,coerce,intern},%
    sensitive=true,%
    morecomment=[l]\#,%
    morestring=[b]',%
    morestring=[b]",%
    morestring=[s]{'''}{'''},
    morestring=[s]{"""}{"""},
    morestring=[s]{r'}{'},
    morestring=[s]{r"}{"},%
    morestring=[s]{r'''}{'''},%
    morestring=[s]{r"""}{"""},%
    morestring=[s]{u'}{'},
    morestring=[s]{u"}{"},%
    morestring=[s]{u'''}{'''},%
    morestring=[s]{u"""}{"""},%
    literate=
    *{+}{{{\color{ipython_purple}+}}}1
    {-}{{{\color{ipython_purple}-}}}1
    {*}{{{\color{ipython_purple}$^\ast$}}}1
    {/}{{{\color{ipython_purple}/}}}1
    {^}{{{\color{ipython_purple}\^{}}}}1
    {?}{{{\color{ipython_purple}?}}}1
    {!}{{{\color{ipython_purple}!}}}1
    {\%}{{{\color{ipython_purple}\%}}}1
    {<}{{{\color{ipython_purple}<}}}1
    {>}{{{\color{ipython_purple}>}}}1
    {|}{{{\color{ipython_purple}|}}}1
    {\&}{{{\color{ipython_purple}\&}}}1
    {~}{{{\color{ipython_purple}~}}}1
    {==}{{{\color{ipython_purple}==}}}2
    {<=}{{{\color{ipython_purple}<=}}}2
    {>=}{{{\color{ipython_purple}>=}}}2
    {+=}{{{+=}}}2
    {-=}{{{-=}}}2
    {*=}{{{$^\ast$=}}}2
    {/=}{{{/=}}}2,
    literate=
    {á}{{\'a}}1 {é}{{\'e}}1 {í}{{\'i}}1 {ó}{{\'o}}1 {ú}{{\'u}}1
    {Á}{{\'A}}1 {É}{{\'E}}1 {Í}{{\'I}}1 {Ó}{{\'O}}1 {Ú}{{\'U}}1
    {à}{{\`a}}1 {è}{{\`e}}1 {ì}{{\`i}}1 {ò}{{\`o}}1 {ù}{{\`u}}1
    {À}{{\`A}}1 {È}{{\'E}}1 {Ì}{{\`I}}1 {Ò}{{\`O}}1 {Ù}{{\`U}}1
    {ä}{{\"a}}1 {ë}{{\"e}}1 {ï}{{\"i}}1 {ö}{{\"o}}1 {ü}{{\"u}}1
    {Ä}{{\"A}}1 {Ë}{{\"E}}1 {Ï}{{\"I}}1 {Ö}{{\"O}}1 {Ü}{{\"U}}1
    {â}{{\^a}}1 {ê}{{\^e}}1 {î}{{\^i}}1 {ô}{{\^o}}1 {û}{{\^u}}1
    {Â}{{\^A}}1 {Ê}{{\^E}}1 {Î}{{\^I}}1 {Ô}{{\^O}}1 {Û}{{\^U}}1
    {œ}{{\oe}}1 {Œ}{{\OE}}1 {æ}{{\ae}}1 {Æ}{{\AE}}1 {ß}{{\ss}}1
    {ç}{{\c c}}1 {Ç}{{\c C}}1 {ø}{{\o}}1 {å}{{\r a}}1 {Å}{{\r A}}1
    {€}{{\EUR}}1 {£}{{\pounds}}1,
    commentstyle=\color{ipython_cyan}\ttfamily,
    stringstyle=\color{ipython_red}\ttfamily,
    keepspaces=true,
    showspaces=false,
    showstringspaces=false,
    rulecolor=\color{ipython_frame},
    frame=single,
    frameround={t}{t}{t}{t},
    framexleftmargin=5mm,
    xleftmargin=2.5em,
    xrightmargin=0.6em,
    numbers=left,
    numberstyle=\tiny\color{halfgray},
    backgroundcolor=\color{ipython_bg},
    basicstyle=\scriptsize\ttfamily,
    keywordstyle=\color{ipython_green}\ttfamily,
}

\newcommand{\im}{{\mathrm{i}}}
\newcommand{\me}{\mathrm{e}}
\newcommand{\ve}{\mathbf{r}}
\newcommand{\Ve}{\mathbf{R}}

\newcommand{\ke}{\mathbf{k}}

\newcommand{\cvo}{\ensuremath{\textrm{CaVO}_3}}%
\newcommand{\sro}{\ensuremath{\textrm{Sr}_2\textrm{RuO}_4}}%
\newcommand{\ceo}{\ensuremath{\textrm{Ce}_2\textrm{O}_3}}%
\newcommand{\phibra}[1]{\ensuremath{\bra{\phi_{{#1} \ke}}}}%
\newcommand{\phiket}[1]{\ensuremath{\ket{\phi_{{#1} \ke}}}}%
\newcommand{\phiketopt}[1]{\ensuremath{\ket{\phi_{{#1} \ke}^{(\text{opt})}}}}%
\newcommand{\wfket}[1]{\ensuremath{\ket{w_{{#1} \ke}}}}%
\newcommand{\done}{\ensuremath{d^1}}
\newcommand{\umit}{\ensuremath{U_{\textrm{\sc mit}}}}
\newcommand{\ttg}{\ensuremath{\textrm{t}_{\textrm{2g}}}}

\usepackage{amssymb}

\definecolor{muted_0}{rgb}{0.282352941176,0.470588235294,0.81568627451}
\definecolor{muted_1}{rgb}{0.933333333333,0.521568627451,0.290196078431}

\newcounter{bla}

\begin{document}

\title{
Charge self-consistent electronic structure calculations with dynamical mean-field theory 
using \acl{QE}, \acl{W90} and \textsc{TRIQS}
}

\author{Sophie Beck}
\email{sbeck@flatironinstitute.org}
\affiliation{Center for Computational Quantum Physics, Flatiron Institute, 162 5th Avenue, New York, NY 10010, USA}
\author{Alexander Hampel}
\affiliation{Center for Computational Quantum Physics, Flatiron Institute, 162 5th Avenue, New York, NY 10010, USA}
\author{Olivier Parcollet}
\affiliation{Center for Computational Quantum Physics, Flatiron Institute, 162 5th Avenue, New York, NY 10010, USA}
\affiliation{Université Paris-Saclay, CNRS, CEA, Institut de physique théorique, 91191, Gif-sur-Yvette, France}
\author{Claude Ederer}
\affiliation{Materials Theory, ETH Z\"u{}rich, Wolfgang-Pauli-Strasse 27, 8093 Z\"u{}rich, Switzerland}
\author{Antoine Georges}
\affiliation{Center for Computational Quantum Physics, Flatiron Institute, 162 5th Avenue, New York, NY 10010, USA}
\affiliation{Collège de France, 11 place Marcelin Berthelot, 75005 Paris, France}
\affiliation{CPHT, CNRS, École Polytechnique, Institut Polytechnique de Paris, Route de Saclay, 91128 Palaiseau, France}
\affiliation{DQMP, Université de Genève, 24 Quai Ernest Ansermet, CH-1211 Genève, Switzerland}

\begin{abstract}
We present a fully \acl{CSC} implementation of \ac{DMFT} combined with \ac{DFT} for electronic structure calculations of materials with strong electronic correlations.
The implementation uses the \acl{QE} package for the \acl{DFT} calculations, the \acl{W90} code for the up-/down-folding and the \textsc{TRIQS} software package for setting up and solving the \ac{DMFT} equations.
All components are available under open source licenses, are MPI-parallelized, fully integrated in the respective packages, and use an hdf5 archive interface to eliminate file parsing.
We show benchmarks for three different systems that demonstrate excellent agreement with  existing \ac{DFT}+\ac{DMFT} implementations in other \emph{ab-initio} electronic structure codes.
\end{abstract}

\maketitle

\begin{acronym}[CT-QMC]
 \acro{CSC}{charge self-consistent}
 \acro{DC}{double counting}
 \acro{DFT}{density functional theory}
 \acro{DMFT}{dynamical mean field theory}
 \acro{FT}{Fourier transform}
 \acro{KS}{Kohn-Sham}
 \acro{MIT}{metal-insulator transition}
 \acro{MLWF}{maximally localized Wannier function}
 \acro{NSCF}{non-self-consistent field}
 \acro{OS}{one-shot}
 \acro{QE}{\textsc{Quantum~ESPRESSO}}
 \acro{SCF}{self-consistent field}
 \acro{TB}{tight-binding}
 \acro{VASP}{``Vienna Ab initio Simulation Package''}
 \acro{W90}{\textsc{Wannier90}}
 \acro{WF}{Wannier function}
\end{acronym}

\acresetall

\section{Introduction}
\label{sec:intro}

\Ac{DMFT} is a conceptual framework and a computational method to deal with strong correlations 
between interacting quantum particles. It can be combined with electronic structure methods, such as \ac{DFT} or the GW method in order to compute and predict the properties of materials with strong electronic correlations (for reviews, see~\cite{held2007, Kotliar:2006}). 
\Ac{DMFT} is the poster child of {\it quantum embedding} methods.
Central to these approaches is the definition of a subset $\mathcal{C}$ of the total Hilbert space, which is spanned by a set of appropriately chosen local orbitals. 
A high-level method (here, \ac{DMFT}) is used for treating many-body effects in the correlated subspace, while the rest of the Hilbert space 
is treated with a simpler method, such as approximations to \ac{DFT} or perturbation theory. 
Any quantity in the full Hilbert space can be projected onto the correlated subspace. When applied to the 
full Green's function, this defines a local projected Green's function which is the key observable on which \ac{DMFT} focuses. 
Correspondingly, the result of the many-body calculation in $\mathcal{C}$ can be embedded in the full Hilbert space by ``upfolding". 

Two types of approaches have been implemented for constructing the \ac{DMFT} equations, which are colloquially referred to by the community as ``Wannier''- or ``projector''-based methods. 

In the former case, a code like \ac{W90}~\cite{Mostofi_et_al:2014} is typically used to formulate a donwfolded Hamiltonian.
In the simplest case this Hamiltonian is even directly formulated in the correlated subspace itself, corresponding to ``maximal downfolding" - in which case this is the only space considered in the calculation. 
In practice, this is what is generally meant by performing \ac{DFT}+\ac{DMFT} calculations using ``Wannier-''orbitals, and it is often the \emph{modus operandi} when using \ac{W90}.
This approach is closer in spirit to model calculations but does not take full advantage of the broader \ac{DFT}+\ac{DMFT} framework. 
 
In more general implementations, the projection/downfolding and embedding/upfolding procedures are performed by defining projectors from the total Hilbert space to the correlated subspace, which involve the overlaps between the local orbitals spanning $\mathcal{C}$ and the basis functions spanning the full space. 
These implementations are colloquially referred to as ``projector-based" and are closer in spirit to what is done in  ``$+U$" extensions of \ac{DFT} (which, incidentally, correspond to solving the \ac{DMFT} embedded problem with a static mean-field approximation). 
It should be noted that, once constructed, the projectors onto $\mathcal{C}$ also yield a set of Wannier functions spanning this subspace, but the user typically does not have direct access to these \acp{WF} in order  to look at dispersions or real-space density as is readily available using \ac{W90}.
While historically the naming convention of ``Wannier''- versus ``projector''-based approach appears reasonable, the implementation presented here requires an update of this terminology. 

In this article, we present a ``projector-based Wannier'' implementation in which the \ac{DFT}+\ac{DMFT} framework is implemented in its general form in a larger Hilbert space (e.g. using a basis set of Kohn-Sham Bloch wavefunctions), while using projector functions computed from \ac{W90} instead of internally in the \ac{DFT} code.

Besides the obvious advantage of benefiting from the flexibility and ongoing developments of \ac{W90}, our implementation also allows to implement full charge self-consistency, which is typically not done when using a maximally downfolded Wannier Hamiltonian. 
Indeed, the  \ac{DFT}+\ac{DMFT} construction relies on a (free-~) energy functional which depends both on the local charge density in the full Hilbert space and on the projected frequency-dependent local Green's function in $\mathcal{C}$. 
Extremalizing this energy functional requires to reach a self-consistent solution for both of these observables. 
Hence, the local charge density is modified by the many-body contributions as compared to a pure \ac{DFT} calculation. 
This is physically important, especially for those materials in which electrons reside in very localized orbitals. 
The simplest approximations to \ac{DFT} such as LDA incorrectly describe these states as itinerant, hence yielding an incorrect charge density and typically a too small value of the equilibrium unit cell volume. 

In practice however, many electronic structure calculations with \ac{DMFT} report results in which the self-consistency over the charge density has not been enforced.
In such calculations, usually called \ac{OS} in the jargon of the field, \ac{DMFT} is only used as a post-processing tool to include many-body effects on top of a converged \ac{DFT} calculation. 
Several codes however are available which combine \ac{DFT} and \ac{DMFT} in order to achieve charge self-consistency, such as implementations based on the Wien2k, VASP, Elk and other electronic structure codes~\cite{aichhorn_dfttools_2016,Schueler_et_al:2018, James:2021, Haule:2010, Amadon:2008,choi_2019}.
All these implementations use projectors constructed internally, rather than Wannier orbitals constructed using \ac{W90}. 

Up to now relatively few systematic studies on the importance of charge self-consistency are available.
While some reports identified systems in which charge self-consistency may be neglected and others in which it plays a crucial role~\cite{Bhandary:2016, Hampel/Beck/Ederer:2020}, little is known about general trends. 
This question is likely to be especially relevant in the context of heterostructures of correlated materials in which charge ordering or interfacial charge transfer take place. 
Thus, with rapidly progressing method developments and rising computational power which allow to treat such problems with multiple correlated atoms within \ac{DMFT}, charge self-consistency becomes an increasingly important question. 

Here, we present an implementation of a fully open-source \ac{CSC} \ac{DFT}+\ac{DMFT} computational framework.
It uses the \ac{QE} package~\cite{Giannozzi_et_al:2009} for the \ac{DFT} calculations, the \ac{W90} code~\cite{Mostofi_et_al:2014} for the generation of the up-/down-folding matrices, and the \textsc{TRIQS/DFTTools} software package~\cite{aichhorn_dfttools_2016,parcollet_triqs_2015} for all the \ac{DMFT}-related routines of up-/down-folding and handling of Green's functions.
All parts of the processes are MPI-parallelized and are thus suited for larger systems and denser $\ke$-point meshes.
The modifications are fully integrated within the software packages \ac{QE} (v.7.0) and \textsc{TRIQS/DFTTools} (v.3.1).
A seamless wrapper routine to manage the workflow between \ac{QE}, \ac{W90} and \textsc{TRIQS} is available in the \textsc{solid\_dmft} package~\cite{solid_dmft}, which enables the user to run fully \ac{CSC} calculations based on a single input file.

The paper is structured as follows:
In Sec.~\ref{sec:theory} we introduce the theoretical framework that is the basis of the implementation.
The workflow describing the usage of the different codes to run \ac{CSC} calculations is outlined in Sec.~\ref{sec:implement}.
Finally, we discuss benchmarks in Sec.~\ref{sec:benchmarks} before concluding in Sec.~\ref{sec:conclusion}.

\section{Theoretical framework}
\label{sec:theory}

The central quantity in \ac{DFT}+\ac{DMFT} calculations is the one-electron lattice Green's function,
\begin{align}\label{eq:lattice_gf}
    \hat G(\ke,\im\omega_n) =& \sum_{\nu\nu'} \, \left[ \im\omega_n+\mu-\hat\epsilon(\ke)-\Delta\hat\Sigma(\ke,\im\omega_n) \right]^{-1}_{\nu\nu'} \Pi_{\nu\nu'}(\ke)\,,\nonumber\\
    =& \sum_{\nu\nu'} \, G_{\nu\nu'}(\ke,\im\omega_n) \Pi_{\nu\nu'}(\ke)\,,
\end{align}
i.e. a function of both the quasi-momentum $\ke$ and of frequency $\omega$, or imaginary Matsubara frequency $\im\omega_n$ as used in the following.
Here, $\nu$ runs over bands of a (Kohn-Sham) Hamiltonian diagonal in momentum space,  i.e.  formulated in terms of Bloch states \phiket{\nu} with corresponding eigenvalues $\epsilon_{\nu\ke}$.
We introduce projectors on these states as
\begin{align}\label{eq:projector}
  \Pi_{\nu\nu'}(\ke) &= \phiket{\nu} \phibra{\nu'}\,,
\end{align}
from which it follows that the Bloch Hamiltonian can be written as $\hat\epsilon(\ke) = \sum_{\nu} \epsilon_{\nu\ke} \Pi_{\nu\nu}(\ke)$.
The chemical potential in eq.~\ref{eq:lattice_gf} is given by $\mu$, and the corrections to the single-particle eigenenergies are represented in terms of the self-energy matrix denoted by $\Delta\hat\Sigma(\ke,\im\omega_n)$ (defined later).
At the initial step of the  \ac{DFT}+\ac{DMFT} iteration,  this term may be set to zero.  

Central to the \ac{DMFT} framework is the definition of a subspace of the total Hilbert space within which strong electronic 
correlations are taken into account.
This subspace is denoted by $\mathcal{C}$, defined more precisely in Sec.~\ref{subsubsec:HS}, and is spanned by a set of local orbitals indexed by $m$. 
We need to define appropriate transformations to perform the transition between the full Hilbert space and the correlated subspace.
This can be achieved by introducing projector functions $P_{m\nu}(\ke)$, which carry out the down-folding, 
i.e. the transformation from the delocalized band-space to the localized orbital-space with indices $\nu$ and $m$, respectively.
The key assumption made in  \ac{DMFT} is that the self-energy can be approximated as a spatially 
local (momentum-independent) matrix {\it when considered in the basis of local orbitals}. 
Hence, the self-energy matrix in the full Hilbert space is obtained as: 
\begin{align}\label{eq:upfolding}
    \Delta\Sigma_{\nu\nu'}(\ke,\im\omega_n) = \sum_{\nu\nu'} P_{\nu m}^{*}(\ke) \Delta\Sigma_{mm'}(\im\omega_n) P_{m'\nu'}(\ke)\,.
\end{align}
in which $\Delta\hat\Sigma$ is the difference between the impurity self-energy and the \ac{DC} correction, 
i.e. $\Delta\hat\Sigma = \hat\Sigma - \hat\Sigma^{\text{DC}}$. 
Note that in this expression, the projector functions $P_{m\nu}(\ke)$ are used to {\it upfold} the self-energy in the correlated subspace to 
the full Hilbert space. 

The local self-energy $\Delta\Sigma_{mm'}(\im\omega_n)$ is calculated within \ac{DMFT} by solving a `quantum impurity 
problem' corresponding to a self-consistent embedding of the correlated atomic shell associated with $\mathcal{C}$~\cite{Georges/Kotliar:1992}. 
This quantum impurity problem is defined by (i) a local hamiltonian of many-body interactions and 
(ii) a frequency-dependent dynamical mean-field (or Weiss field) which specifies the coupling between the embedded atomic shell and 
the self-consistent bath. 
The dynamical mean-field is determined iteratively, and in order to update its value from 
one iteration to the next, one needs to project the full Green's function (\ref{eq:lattice_gf}) onto the local correlated subspace. 
This {\it downfolding} procedure is achieved by using the projectors and by performing a momentum-summation according to: 
\begin{align}\label{eq:downfolding}
    G^{\mathrm{loc}}_{mm'}(\im\omega_n) = \sum_{\ke,\nu\nu'}P_{m\nu}(\ke) G_{\nu\nu'}(\ke,\im\omega_n) P_{\nu'm'}^{*}(\ke)\,.
\end{align}
The Weiss field $\mathcal{\hat G}_0$ is then obtained from the Dyson equation $\mathcal{\hat G}_0^{-1} = \hat\Sigma+\hat G_{\mathrm{loc}}^{-1}$.
In this article, we do not describe the algorithms that are available to solve the impurity problem and refer the reader to extensive reviews of appropriate methods for details~\cite{Gull:2011}.

\begin{figure}
  \centering
  \includegraphics[width=\linewidth]{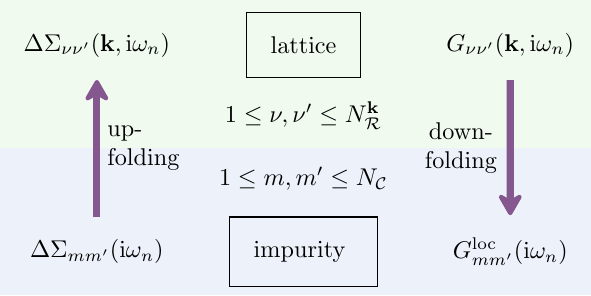}
  \caption{Schematic of up- and down-folding between impurity and lattice space.}
  \label{fig:folding}
\end{figure}

Fig.~\ref{fig:folding} summarizes schematically the upfolding and downfolding procedures between the full Hilbert space 
and the subspace of correlated states. 

\subsection{Up-/Down-folding}

\subsubsection{Hilbert spaces}
\label{subsubsec:HS}

In this section, we specify more precisely the different Hilbert spaces considered in this paper and in our implementation. 
We define four different Hilbert spaces, continually reducing the number of bands/orbitals until we obtain the correlated subspace $\mathcal{C}$ that we aim to treat within \ac{DMFT}.
As starting point we choose a general \ac{KS} Hamiltonian of a non-magnetic \ac{DFT} calculation.
Note that depending on the specifics of the system, some or all Hilbert spaces may be the same.

\begin{figure}
  \centering
  \includegraphics[width=\linewidth]{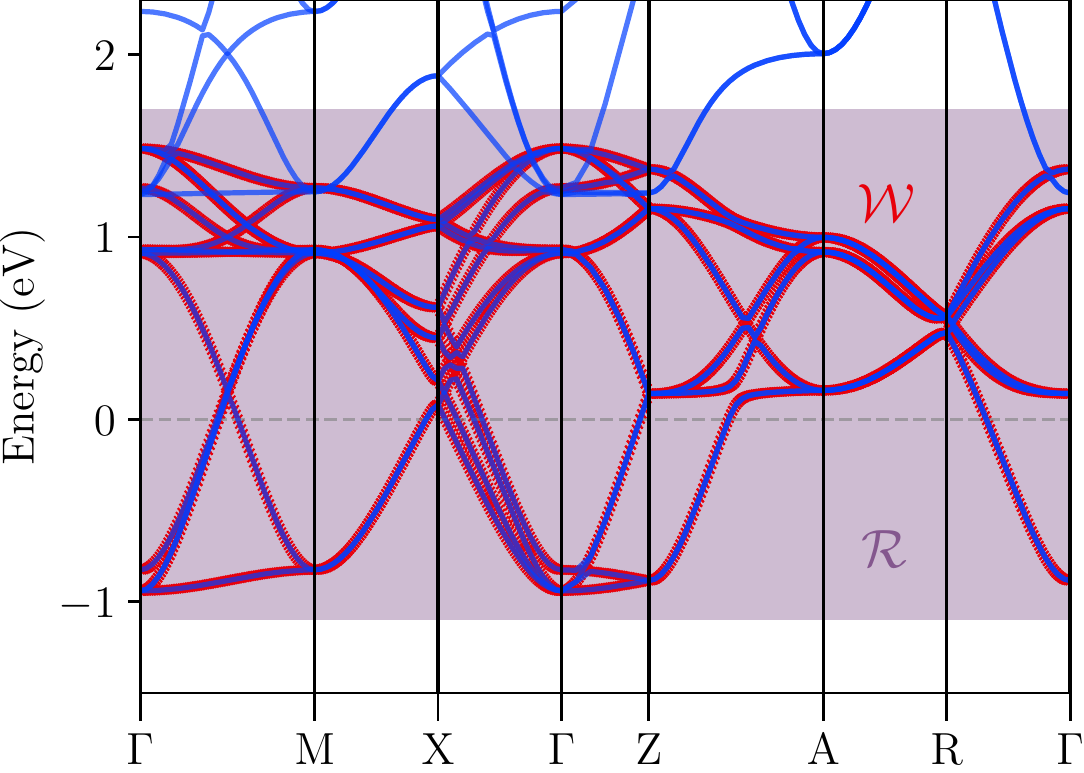}
  \caption{Exemplary definitions of Hilbert spaces described in Sec.~\ref{para:case3} ($N_{\mathcal{B}} > N_{\mathcal{R}}^{\ke} > N_{\mathcal{W}}$) of a $\ke{}$-dependent outer window $\mathcal{R}$ and the optimized and orthonormalized derived Hilbert space $\mathcal{W}$.
  Bands in $\mathcal{B} \ (\mathcal{R})$ and $\mathcal{W}$ are shown in blue and red, respectively.
  We use the cubic perovskite SrVO$_3$ in a $\sqrt{2}\times\sqrt{2}\times 2$ supercell with projections onto an atomic $d$-shell for demonstration purposes.}
  \label{fig:folding_svo}
\end{figure}

We start from the full Bloch space $\mathcal{B}$ associated with the complete basis of \ac{KS} states \phiket{\nu} of dimension \mbox{$N_{\mathcal{B}} \coloneqq$ dim($\mathcal{B}) = \infty$}.
While in theory this could contain infinitely many bands, in practice, it is clear that this space only serves as a hypothetical construct, since the number of \ac{KS} states is finite in any \ac{DFT} calculation.
Yet, a large fraction of \ac{KS} states is described reasonably well within \ac{DFT}, so it is unnecessary to define \acp{WF} for such states, e.g. semicore or other states far away from the Fermi level.
More important is, thus, the reduced \ac{KS} space $\mathcal{R}$, i.e. a subspace of \ac{KS} states limited either via band indices or an energy window (``outer window'' in \textsc{Wannier90}).
We define its dimension as \mbox{$N_{\mathcal{R}}^{\ke} \coloneqq$ dim($\mathcal{R}) < N_{\mathcal{B}}$}, which may carry a $\ke$-dependence in case of entanglement (see Fig.~\ref{fig:folding_svo}).
For this reason, it is evident that such a basis does not constitute a suitable starting point for the definition of a \ac{TB} Hamiltonian.
However, one may construct such a TB Hamiltonian via an optimization/orthonormalization of the Hilbert space, implemented in \textsc{Wannier90}, details of which can be found in the appropriate literature~\cite{Marzari_et_al:2012}.
This allows to define a Hilbert space $\mathcal{W}$, containing \mbox{$N_{\mathcal{W}} \coloneqq$ dim($\mathcal{W}) \leq \min(N_{\mathcal{R}})$} optimized and orthonormalized Bloch states \phiketopt{\mu} with band indices $\mu, \mu'$.
A straight-forward gauge transformation to orbital space then defines \acp{WF} \wfket{\alpha} with orbital indices $\alpha, \alpha'$, see Sec.~\ref{subsubsec:proj}.
Finally, one may select a subset of $m$ \acp{WF} from this complete basis as the ``correlated subspace'' $\mathcal{C}$, i.e. $N_{\mathcal{C}} \coloneqq$ dim($\mathcal{C}) \leq N_{\mathcal{W}}$.

It now becomes clear that, generally speaking, the up- and down-folding occurs between the Hilbert spaces $\mathcal{R}$ and $\mathcal{C}$.
Therefore, the main task of the following manuscript is to find the correct projector functions $\hat P(\ke)$ that allow up- and down-folding between the orbital space $\mathcal{C}$ (and more generally $\mathcal{W}$) and the space $\mathcal{R}$ of Bloch states, illustrated in Fig.~\ref{fig:folding}.
In the following the terms and notations referring to the original Bloch states denote the respective objects in $\mathcal{R}$ and are labelled with index $\nu$.
Equivalently, objects in $\mathcal{W}$ will be labelled by $\mu$ if considered in the diagonal band gauge, and $\alpha$ if formulated in the orbital gauge, where $\alpha = \{1,...,N_{\mathcal{C}},...N_{\mathcal{W}}\}$.
Finally, correlated orbitals with index $m$ belong to $\mathcal{C}$.
Each correlated orbital $m$ is associated with an atomic site $i$, meaning that the orbital space $\mathcal{C}$ can be spanned by multiple sites.
Note that in general the construction of projector functions $P_{\alpha\nu}(\ke)$ for $\alpha > N_{\mathcal{C}}$ is not strictly necessary as it is used only for post-processing purposes, but not in the \ac{DMFT} equations. 

\subsubsection{Projector functions}
\label{subsubsec:proj}

In the following, we consider three cases that capture the different scenarios for (hypothetical and) usual applications.

\paragraph{Case 1 - infinite set of bands: $N_{\mathcal{B}} = N_{\mathcal{R}} = N_{\mathcal{W}}$}
\label{para:case1}

First, we construct a set of $N_{\mathcal{W}}$ \acp{WF} from the entire set of $N_{\mathcal{B}}$ \ac{KS} states.
While the case of an infinite set of bands is clearly hypothetical, we use this paragraph to introduce the basic quantities and concepts.
The \acp{WF} \wfket{\alpha} are obtained simply via change of basis as
\begin{align}\label{eq:wannier}
    \wfket{\alpha} = \sum_{\nu=1}^{N_{\mathcal{B}}} U_{\nu\alpha}^{*(\ke)} \phiket{\nu}\, ,
\end{align}
where we introduced unitary matrices $U_{\alpha\nu}^{(\ke)}$ which define the gauge.
The \acp{WF} may alternatively be expressed in real space as centered at lattice site $\Ve{}_i$ via a \ac{FT}, i.e.
\begin{align}
    \ket{w_{\alpha i}} = \frac{V}{(2\pi)^3} \int_{\text{BZ}} d \ke \me^{-\im \ke\Ve_i} \wfket{\alpha}\,.
\end{align}
Since, due to the gauge freedom, the $U_{\alpha\nu}^{(\ke)}$ are not uniquely defined, different definitions of \acp{WF} can be found in the literature.
While here we use the spread minimization as implemented in \ac{W90}, an alternative is given in the typical ``projector''-type approach by using code-internal routines to determine the gauge.
The advantage of \acp{MLWF} in \ac{W90} thus lies not only in obtaining a unique solution with ideal localization properties, but also in the flexibility given by a \ac{DFT}-code-independent approach.
Note that the following formalism is general, i.e. remains valid for other definitions of the gauge freedom.
The spread $\Omega$ can be calculated from the \acp{WF} at the respective lattice sites in the unit cell ($\Ve_i = 0$) as:
\begin{align}
    \Omega = \sum_{\alpha}\left[ \braket{w_{\alpha 0}(\ve)|\ve^2|w_{\alpha 0}(\ve)} - \braket{w_{\alpha 0}(\ve)|\ve|w_{\alpha 0}(\ve)}^2 \right].
\end{align}
This can further be divided into a gauge invariant term $\Omega_{\text{I}}$ and a gauge dependent term $\widetilde\Omega$.
The latter is used in the Marzari-Vanderbilt (MV) method~\cite{Marzari/Vanderbilt:1997} to obtain \acp{MLWF}.

The projector functions $\hat P(\ke)$ introduced in eq.~\ref{eq:downfolding} can be readily identified as the unitary matrix, i.e. the transformation from the Bloch to orbital gauge:
\begin{align}\label{eq:p1}
    P_{\alpha\nu}(\ke) \equiv U_{\alpha\nu}^{(\ke)}\,.
\end{align}
For the initial guess of the projector function so-called trial localized orbitals $\ket{g_{\alpha}}$ are used in \textsc{Wannier90}, e.g. atomic-like states.
The initial overlap to the \ac{KS} states is defined as
\begin{align}
\label{eq:init}
    \widetilde U_{\alpha\nu}(\ke) =  \braket{g_{\alpha}|\phi_{\nu\ke}}\,.
\end{align}
Note that unless the trial localized orbitals $\ket{g_{\alpha}}$ are fully contained within $\mathcal{B}$ (i.e. if $N_{\mathcal{B}} = \infty$), the projector functions need to orthonormalized (see the following paragraph, i.e. case 2).
During the spread minimization the $U_{\alpha\nu}^{(\ke)}$ are then optimized to yield the resulting \ac{MLWF}.

\paragraph{Case 2 - isolated set of bands: $N_{\mathcal{B}} > N_{\mathcal{R}} = N_{\mathcal{W}}$}
\label{para:case2}

We now assume that only a subset $N_{\mathcal{R}}$ of the total number of \ac{KS} states contributes to the construction of the Wannier basis.
We further assume that it constitutes an isolated set of states, meaning that $N_{\mathcal{R}}^{\ke}=N_{\mathcal{R}}$ is $\ke{}$-independent.
In particular this means that no \ac{KS} states \phiket{\nu} from outside this window cross target bands at any $\ke$.
In this case, $\mathcal{B}$ trivially divides into two orthogonal subspaces, $\mathcal{B}=\mathcal{R} \oplus \mathcal{R}^{\perp}$, where each \ac{KS} Bloch function is either fully contained in $\mathcal{R}$ or in $\mathcal{R}^{\perp}$, so that $\phiket{\nu}$ with $\nu \in \mathcal{R}$ form a complete and orthogonal basis in $\mathcal{R}$.
However, if the projectors are initialized as in eq.~\ref{eq:init}, they need to be orthonormalized, since in general the trial localized orbitals $\ket{g_{\alpha}}$ are not fully contained in $\mathcal{R}$.
Thus, we first define non-orthogonal $\ket{\widetilde w_{\alpha\ke}}$ from the \ac{KS} states within $N_{\mathcal{R}}$ (in this paragraph we will omit the superscript $\ke{}$ in $N_{\mathcal{R}}^{\ke}$, since we assume an isolated set of bands), i.e.
\begin{align}\label{eq:non-orth}
    \ket{\widetilde w_{\alpha\ke}} = \sum_{\nu=1}^{N_{\mathcal{R}}} \widetilde U_{\nu\alpha}^{*(\ke)} \phiket{\nu}\,.
\end{align}
To orthogonalize we calculate the overlap as
\begin{align}
    O_{\alpha\alpha'}(\ke) \coloneqq \braket{\widetilde w_{\alpha\ke} | \widetilde w_{\alpha'\ke}} = \sum_{\nu=1}^{N_{\mathcal{R}}} \widetilde U_{\alpha\nu}^{(\ke)} \widetilde U_{\nu\alpha'}^{*(\ke)} \,,
\end{align}
and, for a simplified notation we further define $S_{\alpha\alpha'}(\ke) = \{O(\ke)^{-1/2}\}_{\alpha\alpha'}$.
Following this, the orthonormal set of \acp{WF} is given as
\begin{align}
    \wfket{\alpha} =& \sum_{\alpha'=1}^{N_{\mathcal{W}}} S_{\alpha\alpha'}(\ke) \ket{\widetilde w_{\alpha'\ke}}\,%
    = \sum_{\alpha'=1}^{N_{\mathcal{W}}} \sum_{\nu=1}^{N_{\mathcal{R}}} S_{\alpha\alpha'}(\ke) \widetilde U_{\nu\alpha'}^{*(\ke)} \phiket{\nu}\, \nonumber\\%
    =& \sum_{\nu=1}^{N_{\mathcal{R}}} U_{\nu\alpha}^{*(\ke)} \phiket{\nu}\,,
\end{align}
with 
\begin{align}\label{eq:p2}
    U_{\nu\alpha}^{*(\ke)} \coloneqq \sum_{\alpha'=1}^{N_{\mathcal{W}}} S_{\alpha\alpha'}(\ke) \widetilde U_{\nu\alpha'}^{*(\ke)}\,.
\end{align}
The final orthonormal (and potentially spread-minimized) set of unitary projector functions is then given as $P_{\alpha\nu}(\ke) \equiv U_{\alpha\nu}^{(\ke)}$ as above.

Finally, we note that in this specific case of an isolated set of bands ($N_{\mathcal{R}}=N_{\mathcal{W}}$) it is possible to use the set of Wannier states in $\mathcal{W}$ as a basis set during the \ac{DMFT} cycle instead of using the \ac{KS} states (in the following paragraph we use superscripts \ac{TB} and \ac{KS} for clarity).
The advantage of the Wannier basis is that it has a more direct physical interpretation in terms of atomic orbitals. 
This is possible because the same set of bands is included in the chosen window at all $\ke$-points, so that the one-particle \ac{KS} Hamiltonian can be unitarily rotated to this \ac{TB} basis:
\begin{align}\label{eq:downfolding_hloc}
    \epsilon^{\mathrm{TB}}_{\alpha\alpha'}(\ke)= \sum_{\nu} P_{\alpha\nu}(\ke) \epsilon^{\mathrm{KS}}_{\nu}(\ke) P_{\nu \alpha'}^{*}(\ke)\,,
\end{align}
The ``effective projector functions" used in the code can be expressed in this basis and simply become \mbox{$P^{\mathrm{TB}}_{m\alpha}(\ke)=\delta_{m,\alpha}$}.
The upfolding of the self-energy (eq.~\ref{eq:upfolding}) becomes trivial in this basis:
it simply amounts to constructing the self-energy $\Delta\hat{\Sigma}_{\alpha\alpha'}$ as a matrix with a block equal to $\Delta\hat{\Sigma}_{mm'}$ if $\alpha,\alpha' \leq N_{\mathcal{C}}$ and $0$ otherwise.
The lattice Green's function in the Wannier (\ac{TB}) basis and projected Green's function in $\mathcal{C}$ simply read: 
\begin{eqnarray}
\label{eq:lattice_gf_Wannier}
G(\ke,\im\omega_n)_{\alpha\alpha'}&=&\left[ \im\omega_n+\mu-\hat\epsilon(\ke)-\Delta\hat\Sigma(\ke,\im\omega_n) \right]^{-1}_{\alpha\alpha'}\nonumber \\
G^{\mathrm{loc}}_{mm'}(\im\omega_n)&=&\sum_{\ke}G(\ke,\im\omega_n)_{mm'} 
\end{eqnarray}
and the DMFT self-consistency loop is implemented as usual as $\mathcal{\hat G}_0^{-1} = \hat\Sigma+\hat G_{\mathrm{loc}}^{-1}$.
For practical purposes we define here the diagonalized Wannier Hamiltonian
\begin{align}\label{eq:TBHamiltonian}
    \hat\epsilon^{\mathrm{TB}}_{\mu}(\ke) = \sum_{\mu} \epsilon_{\mu\ke} \Pi_{\mu\mu}(\ke)\,,
\end{align}
which is identical to $\epsilon^{\mathrm{KS}}_{\nu}(\ke)$ only in the case of an isolated set of bands.

We will use the term ``Wannier mode" when the \ac{DMFT} equations are implemented in this manner using the Wannier basis, while the term ``Band mode" is used to designate the more general implementation in the basis of \ac{KS} Bloch states. 
Our implementation allows the user to choose one mode or the other. 
We emphasize however that the ``Wannier mode" is only correct for an isolated set of bands and should be used only in this case, i.e. if the projector functions $P_{\alpha\nu}(\ke)$ are unitary (square) matrices, as outlined in Appendix \ref{sec:appendixA}. 
The use of Wannier functions constructed with \ac{W90} to define the projectors, while implementing the \ac{DMFT} cycle in a general manner in the \ac{KS} basis is one of the novel aspects of our implementation.

\paragraph{Case 3 - entangled bands: $N_{\mathcal{B}} > N_{\mathcal{R}}^{\ke} > N_{\mathcal{W}}$}
\label{para:case3}

Finally, one may construct \acp{WF} from a set of bands $N_{\mathcal{R}}^{\ke}$ that is not isolated, i.e. it contains \ac{KS} states that are undesired in the construction of the \acp{WF}.
This means that within a specified energy window (``outer window'') and/or within a specified range of band indices in $\mathcal{B}$ it may contain a ($\ke$-dependent) number of additional \ac{KS} states, illustrated in Fig.~\ref{fig:folding_svo}.
The additional states are typically of a different orbital character and thus undesirable when selecting the correlated subspace, but can be excluded via a disentanglement procedure.
To this end we first define disentanglement matrices $U_{\mu\nu}^{\text{dis}(\ke)}$, which are generally not square but rectangular matrices of dimension \mbox{$N_{\mathcal{W}} \times N_{\mathcal{R}}^{\ke}$}, used to optimize the cell-periodic part of the \ac{KS} states within $\mathcal{R}$:
\begin{align}
    \phiketopt{\mu} = \sum_{\nu=1}^{N_{\mathcal{R}}^{\ke}} U_{\mu\nu}^{\text{dis}(\ke)} \phiket{\nu}\,.
\end{align}
This can be achieved by minimizing the gauge invariant term of the spread, $\Omega_{\text{I}}$, which ensures $\ke$-point connectivity, or ``global smoothness of connection'' in the optimized states~\cite{Souza/Marzari/Vanderbilt:2001}.
This leaves us with $N_{\mathcal{W}}$ optimized \ac{KS} states \phiketopt{\mu}, which can further be spread-minimized
as outlined in the previous paragraph:
\begin{align}
    \wfket{\alpha} =& \sum_{\mu=1}^{N_{\mathcal{W}}} U_{\mu\alpha}^{*(\ke)} \phiketopt{\mu}
    = \sum_{\mu=1}^{N_{\mathcal{W}}} \sum_{\nu=1}^{N_{\mathcal{R}}^{\ke}} U_{\mu\alpha}^{*(\ke)} U_{\mu\nu}^{\text{dis}(\ke)} \phiket{\nu}\,.
\end{align}
Following this, the projector functions used for the up-/down-folding are then given by
\begin{align}\label{eq:p3}
    P_{\alpha\nu}(\ke) \coloneqq U_{\mu\alpha}^{*(\ke)} U_{\mu\nu}^{\text{dis}(\ke)}\,.
\end{align}

Note that in this case the projector functions may be rectangular (since $N_{\mathcal{R}}^{\ke}\neq N_{\mathcal{W}}$), i.e. need not necessarily be unitary (i.e. satisfy $\hat P\hat P^{\dagger} = \mathbb{\hat 1} \neq \hat P^{\dagger}\hat P$).
This means that the down-folding in eq.~\ref{eq:downfolding} from Bloch to Wannier basis is not invertible, that is the reverse process of up-folding does not recover the full \ac{TB} Hamiltonian and lattice Green's function.
In these cases a formulation of the \ac{DMFT} equations in an orbital basis as outlined in the previous paragraph (case 2) and in the Appendix \ref{sec:appendixA} becomes an approximation to the \ac{DMFT} formulation in the \ac{KS} basis.

We close the discussion of the different subspaces by remarking that the case 2 leads to projector functions $\hat P$, which are very similar to those obtained in the typical ''projector''-based approach for an isolated set of bands~\cite{Hampel/Beck/Ederer:2020}.
In contrast, for cases 1 and 3 discussed the projector functions may differ more drastically due to the absence of a disentanglement procedure to match the underlying \ac{DFT} band structure in the typical ''projector''-based approach.
In other words, the projector functions described here result not only from a raw projection on atomic orbitals, but also incorporate some of the complexity of orbitals in a real solid by respecting the underlying band structure such as ligand bonds, while at the same time optimizing the localization condition.

\subsection{Charge self-consistency and total energy}

This section provides the most important equations required for \ac{CSC} calculations, closely following 
Refs.~\cite{Lechermann:2006,Lechermann_et_al:2018}.
The interacting charge density at inverse temperature $\beta$ is given by
\begin{align}\label{eq:chargedensity}
    \rho(\ve) = \frac{1}{\beta} \sum_{n,\ke} \bra{\ve} \hat G(\ke,\im\omega_n) \ket{\ve} \equiv \rho^{\text{KS}}(\ve) + \Delta\rho(\ve)\,.
\end{align}
Here, the \ac{KS} charge density is defined as
\begin{align}
  \rho^{\text{KS}}(\ve) &= \sum_{\ke{}}  \sum_{\nu=1}^{N_{\mathcal{B}}} f_{\nu\ke}^{\text{KS}} \braket{\ve|\Pi_{\nu\nu}(\ke)|\ve}\,,
\end{align}
with the KS occupation numbers $f_{\nu\ke}^{\text{KS}}$ and the KS Fermi level $\mu^{\text{KS}}$ chosen such that the correct total charge is obtained within the \ac{DFT} calculation.
Note that for each iteration $\rho^{\text{KS}}(\ve)$ represents the non-interacting contribution to the physical charge density $\rho(\ve)$, and thus is not identical to the true \ac{DFT} ground state density $\rho^{\text{KS}}_{0}(\ve)$ (only in the first iteration).
In order to obtain the correction to the non-interacting charge density one needs to calculate the corrected occupations by subtracting the non-interacting part as follows:
\begin{align}
    \Delta \rho(\ve) =& \frac{1}{\beta}\sum_{n,\ke} \bra{\ve} \hat G(\ke,\im\omega_n) - \hat G^{\text{KS}}(\ke,\im\omega_n) \ket{\ve}\,,\nonumber\\
    =& \frac{1}{\beta} \sum_{n,\ke} \bra{\ve} \hat G^{\text{KS}}(\ke,\im\omega_n) [\Delta\hat\Sigma(\ke,\im\omega_n) -\Delta\hat\mu]\nonumber\\
    & \times \hat G(\ke,\im\omega_n) \ket{\ve}\,,
\end{align}
where $\Delta\hat\mu = (\mu-\mu^{\text{KS}})\mathbb{\hat 1}$ is the difference in chemical potentials.
The second line in the equation above is derived using the definition $(\hat G^{\text{KS}})^{-1} = (\im \omega_n + \mu^{\text{KS}})\mathbb{\hat 1}  - \hat \epsilon$ and $\hat{G}-\hat{G}_{KS} = \hat{G}_{KS} [\hat{G}_{KS}^{-1}-\hat{G}^{-1}]\hat{G}$.
Note that this expression is better behaved computationally, since the right-hand side decays faster with frequency than the slow decay $\sim 1/i\omega$ of individual Green's functions. 
We now define the correction to the ground state charge density as
\begin{align}
    \Delta \rho(\ve) =&  \sum_{\ke} \bra{\ve} \Delta\hat N(\ke) \ket{\ve}\,,
\end{align}
where the charge density correction $\Delta \hat N(\ke)$ is defined as
\begin{align}\label{eq:hatdeltaN}
    \Delta \hat N(\ke) = \sum_{\nu,\nu'=1}^{N_{\mathcal{R}}^{\ke}} \Delta N_{\nu\nu'} (\ke) \Pi_{\nu\nu'}(\ke)\,,
\end{align}
with the matrix elements given by
\begin{align}\label{eq:deltaN}
    \Delta N_{\nu\nu'}(\ke) \equiv& \frac{1}{\beta} \sum_{n} \phibra{\nu} \hat G^{\text{KS}}(\ke,\im\omega_n) [\Delta\hat\Sigma(\ke,\im\omega_n) \nonumber\\
    & -\Delta\hat\mu] \hat G(\ke,\im\omega_n) \phiket{\nu'}\,,\nonumber\\
    =& \frac{1}{\beta} \sum_n \sum_{\nu'',\nu'''=1}^{N_{\mathcal{R}}^{\ke}} G_{\nu\nu''}^{\text{KS}}(\ke,\im\omega_n) [\Delta\Sigma_{\nu''\nu'''}(\ke,\im\omega_n) \nonumber\\
    &  -\Delta\mu\delta_{\nu''\nu'''}] G_{\nu'''\nu'}(\ke,\im\omega_n)\,.
\end{align}
Note that in order to maintain charge neutrality the trace over the occupation corrections must amount to zero, i.e. the changes in the occupations do not change the total charge:
\begin{align}
    \sum_{\ke}\sum_{\nu=1}^{N_{\mathcal{R}}^{\ke}}\Delta N_{\nu\nu}(\ke) = 0\,.
\end{align}
With these definitions the total interacting charge density in eq.~\ref{eq:chargedensity} can be recast into
\begin{align}
    \rho(\ve) ={}&{} \rho^{\text{KS}}(\ve) + \Delta\rho(\ve)\,,\nonumber\\%
    =& \sum_{\ke} \sum_{\nu,\nu'=1}^{N_{\mathcal{B}}} \braket{\ve | \Pi_{\nu\nu'}(\ke) | \ve} \left[ f_{\nu\ke}^{\text{KS}} \delta_{\nu\nu'} + \Delta N_{\nu\nu'}(\ke) \right]\,.
\end{align}
For practical purposes, it is advisable to absorb the non-diagonal elements by orthonormalizing the \ac{KS} states via a unitary transformation $\hat V \in \mathcal{R}$~\cite{Schueler_et_al:2018}, obtained as
\begin{align}\label{eq:diagonalize}
    \Delta \hat N(\ke) = \hat V \hat D(\ke) \hat V^{-1}\,.
\end{align}
where $\hat D(\ke)$ is diagonal and contains the orthonormalized occupation updates.
After embedding $\hat V$ and $\Delta\hat N$ as block-matrixes into identity matrices in $\mathcal{B}$ we obtain updated \ac{KS} states as
\begin{align}\label{eq:rotate}
    \ket{\widetilde\phi_{\nu\ke}} \coloneqq \sum_{\nu'=1}^{N_{\mathcal{B}}} V_{\nu\nu'} \phiket{\nu'}\,.
\end{align}
Finally, the charge density is given as
\begin{align}\label{eq:updated_cd}
    \rho(\ve) =& \sum_{\ke} \sum_{\nu=1}^{N_{\mathcal{B}}} \widetilde f_{\nu\ke} |\widetilde\phi_{\nu\ke}(\ve)|^2\,,
\end{align}
where the renormalized occupation numbers are defined by
\begin{align}
    \widetilde f_{\nu\ke} \delta_{\nu\nu'} = \sum_{\nu'',\nu'''=1}^{N_{\mathcal{B}}} V_{\nu\nu''} \left[ f_{\nu''\ke}^{\text{KS}} \delta_{\nu''\nu'''} + \Delta N_{\nu''\nu'''}(\ke) \right] V_{\nu'''\nu'}^*\,.
\end{align}

The free energy can be derived from the full \ac{DFT}+\ac{DMFT} functional as the stationary point with respect to the impurity Green's function $\hat G^{\mathcal{C}}$~\cite{Kotliar:2006,Haule:2015_free_energy}.
Here, we are particularly interested in the total energy of the system to draw a comparison with \ac{DFT}, which is given by the limit $T \rightarrow 0$~K 
of the free energy~\cite{Amadon:2006,Pourovskii_et_al:2007} as:
\begin{align}
  E ={}& E_{\sc\mathrm DFT}[\rho] + \sum_{\ke{}} \sum_{\nu=1}^{N_{\mathcal{R}}^{\ke}} \Delta N_{\nu\nu}(\ke) \epsilon_{\nu\ke}\nonumber\\
  & + E_{\mathrm{int}}[\hat G^{\mathcal{C}}] - E_{\sc\mathrm DC}[\hat G^{\mathcal{C}}]\,.
\end{align}
Here, the first and second term correspond to the \ac{DFT} energy and the \ac{DMFT}-derived kinetic energy corrections, respectively.
The second line represents the impurities interaction energy minus the double counting correction in the usual formulation.

\section{Implementation notes}
\label{sec:implement}

\subsection{Workflow}
\label{subsec:workflow}

%
\begin{figure}
  \centering
  \includegraphics[width=\linewidth]{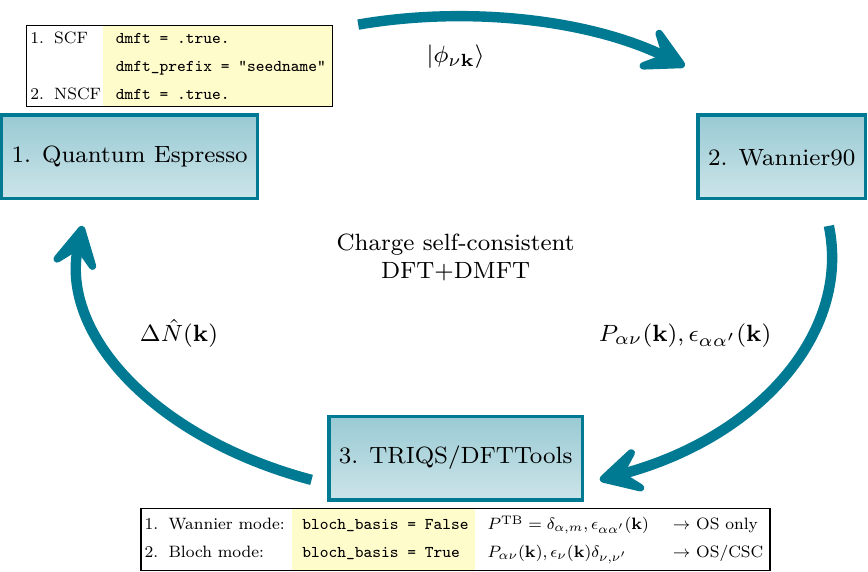}
  \caption{
  Workflow of a fully charge self-consistent \ac{DFT}+\ac{DMFT} calculation (details explained in Sec.~\ref{sec:implement}).
  }
  \label{fig:workflow}
\end{figure}
The workflow of the \ac{CSC} calculation requires three software components, illustrated in Fig.~\ref{fig:workflow} and described in the following.
All additions are fully integrated in the respective open-source software packages.

\subsubsection{\ac{QE} - ground state preparation}
The ground state density $\rho_0^{\text{KS}}(\ve)$ is determined as usual in a \ac{SCF} calculation.
Following this, the user generates wavefunctions $\phiket{\nu}$ on a homogeneous $\ke{}$-point grid in an \ac{NSCF} calculation as usual to interface with \ac{W90}.
Given the \texttt{dmft = .true.} keyword in the \ac{NSCF} system card (see Sec.~\ref{subsubsec:QE}), \ac{QE} will exit, but write the current wavefunctions and corresponding occupations to an hdf5 archive to keep a snapshot of the current state.
This effectively allows to continue the calculation with updated occupations and wavefunctions in the next \ac{DFT} step to obtain the next charge density as described in Sec.~\ref{subsubsec:QE}.
Note that the \textsc{TRIQS/DFTTools} converter expects the output named as \texttt{seedname.nscf.out} with \texttt{verbosity = 'high'} in the control card.

\subsubsection{\ac{W90}}
\label{subsubsec:w90}
Subsequently, \ac{W90} is called with the same \texttt{seedname}, where the flags \texttt{write\_hr} and \texttt{write\_u\_matrices} must be switched on to explicitly write the Hamiltonian $\hat\epsilon_{\alpha\alpha'}$ and matrices $\hat U$ to file.
This will create all necessary output files for the \textsc{TRIQS/DFTTools} converter.

\subsubsection{\textsc{TRIQS}}
\label{subsubsec:triqs}
The projector functions, together with the \ac{KS} eigenvalues and other relevant information is read and written into hdf5 format (\texttt{seedname.h5}) using the Wannier90Converter in \textsc{TRIQS/DFTTools}.
With an appropriate configuration file \texttt{seedname.inp}, which specifies the number of wannier functions, desired $\ke{}$-point grid, and the orbital character, the converter is run with a simple python script:
\begin{lstlisting}[language=iPython]
from triqs_dft_tools.converters import (
                            Wannier90Converter)

converter = Wannier90Converter(seedname='seedname', 
                               bloch_basis=True)
converter.convert_dft_input()
\end{lstlisting}
Importantly, the flag \texttt{bloch\_basis} must be switched on for \ac{CSC} calculations, determining whether the \ac{DMFT} equations are formulated in the Bloch (\texttt{True}) or Wannier basis (\texttt{False}), as outlined in Sec.~\ref{subsubsec:HS} and illustrated in Fig.~\ref{fig:workflow}.
By default the converter leverages the fact that the single-site impurity problems are block-diagonal in the site index $i$, meaning that they decouple and can be solved independently, but in principle they can also be set up as a cluster with off-diagonal elements.

After a number of self-consistent \ac{DMFT} iterations the charge density correction (see eq.~\ref{eq:deltaN}) is computed and stored directly in the existing \texttt{seedname.h5} archive.
This is handled again in \textsc{TRIQS/DFTTools}, using the SumkDFT class.
This class provides lattice Green's function operations and for a detailed overview we refer the reader to Ref.~\cite{aichhorn_dfttools:2016}. The charge density correction is calculated by upfolding the impurity self-energy into the lattice Green's function (see Eq.~\ref{eq:upfolding}), determining the chemical potential, and then calling the function \texttt{calc\_density\_correction} provided within the SumkDFT class.
This marks the endpoint of the \ac{DMFT} cycle.

\subsubsection{\ac{QE} - continued}
\label{subsubsec:QE}

At this point \ac{QE} is restarted in the \ac{SCF} mode at the checkpoint whose creation was triggered at the end of the previous \ac{NSCF} calculation with corresponding wavefunctions and occupations.
Importantly, \ac{QE} skips the determination of new wavefunctions, but computes the updated \ac{DFT}+\ac{DMFT} charge density (Eq.~\ref{eq:updated_cd}) with the following key input parameters given:
\begin{lstlisting}[language=Fortran]
&system
   restart_mode     = 'restart', !SCF only
   dmft             = .true.,
   dmft_prefix      = 'seedname', !SCF only
   nosym            = .true.,
/
&electrons                  
   electron_maxstep = 1, !SCF only
   mixing_beta      = 0.3, !SCF only; 
/
\end{lstlisting}
In particular, the charge density corrections from \ac{DMFT} are read from the \texttt{seedname.h5} archive, and are used to obtain the unitary matrices $\hat V$ (eq.~\ref{eq:diagonalize}), which diagonalize the new occupations and consequently rotate the \ac{KS} basis as defined in eq.~\ref{eq:rotate}.
Based on these new states, the code computes the updated charge density $\rho(\ve)$ (eq.~\ref{eq:updated_cd}), which corresponds to the physical \ac{DFT}+\ac{DMFT} charge density as defined in eq.~\ref{eq:chargedensity}.
\ac{QE} is stopped (\texttt{electron\_maxstep = 1}), implying that no \ac{DFT} self-consistency is reached.
Instead, an \ac{NSCF} calculation follows as described above, computing from $\rho(\ve)$ a new Hamiltonian and corresponding single-particle states $\ket{\phi_{\nu\ke}^{(1)}}$, where the superscript $(1)$ labels the next generation of \ac{KS} states to be downfolded.
The procedure described in Secs.~\ref{subsubsec:w90}, \ref{subsubsec:triqs} and \ref{subsubsec:QE} is repeated until convergence.
Importantly, no information is written to, or read from, text files, in contrast to currently existing \ac{DFT}+\ac{DMFT} interfaces, reducing I/O operations, making the procedure less error-prone, and eliminating the need of elaborate file parsing.

\subsection{Installation and parallelization remarks}

As outlined above the interface consists of two new software parts.
\begin{enumerate}
    \item[1.]
    The first part of the interface is directly implemented in \ac{QE}, starting with version 7.0.
    The compilation requires hdf5 support, since the data exchange between codes is handled on the level of hdf5 archives.
    \item[2.]
    The second part is implemented in \textsc{TRIQS/DFTTools}, starting with version 3.1, which handles the actual conversion from the \ac{W90} output (see Fig.~\ref{fig:workflow} right side).
    \textsc{TRIQS/DFTTools} is publicly available on GitHub at \href{https://github.com/triqs/dft_tools}{github.com/triqs/dft\_tools} and contains all the functionality discussed here.
    \textsc{TRIQS/DFTTools} is installed like any other \textsc{TRIQS} application.
    For explicit instructions see the documentation of \textsc{TRIQS/DFTTools}, which also contains an extensive tutorial on how to use the converter including example input files\footnote{\href{https://triqs.github.io/dft_tools/unstable/guide/conv_W90.html}{triqs.github.io/dft\_tools/unstable/guide/conv\_W90}}.
\end{enumerate}
The \ac{W90} code does not require any modifications, and can be acquired via \href{http:\\www.wannier90.org}{wannier90.org}.  

As outlined above the converter leverages \ac{MPI} for parallelization over $\ke$-points.
This also holds for all other components described above implemented in \ac{QE} and \textsc{TRIQS}, making it scaling well with number of $\ke$-points.
Hence, the converter should be called as \texttt{mpirun python w90\_converter.py}. 

A wrapper routine handling all parts of this workflow including a tutorial is available in the \textsc{solid\_dmft} package~\cite{solid_dmft}, which is publicly available on github \href{https://github.com/flatironinstitute/solid_dmft}{github.com/flatironinstitute/solid\_dmft} and is installed like any other \textsc{TRIQS} application.

\section{Benchmarks}
\label{sec:benchmarks}

The results presented in this section are obtained using the implementation described above.
The implementation is benchmarked for two systems where \ac{CSC} was shown to affect the materials properties.
The first example concerns the strain-induced orbital polarization in \cvo{}, while the second benchmark reports the change in equilibrium volume in \ceo{}. 
We also report on the example of \sro{} where, in contrast, \ac{CSC} is found to play a minor role with no changes in quasiparticle mass renormalizations.

\paragraph{Orbital polarization in \cvo{}}
\label{para:example_cvo}

\begin{figure}
  \centering
  \includegraphics[width=\linewidth]{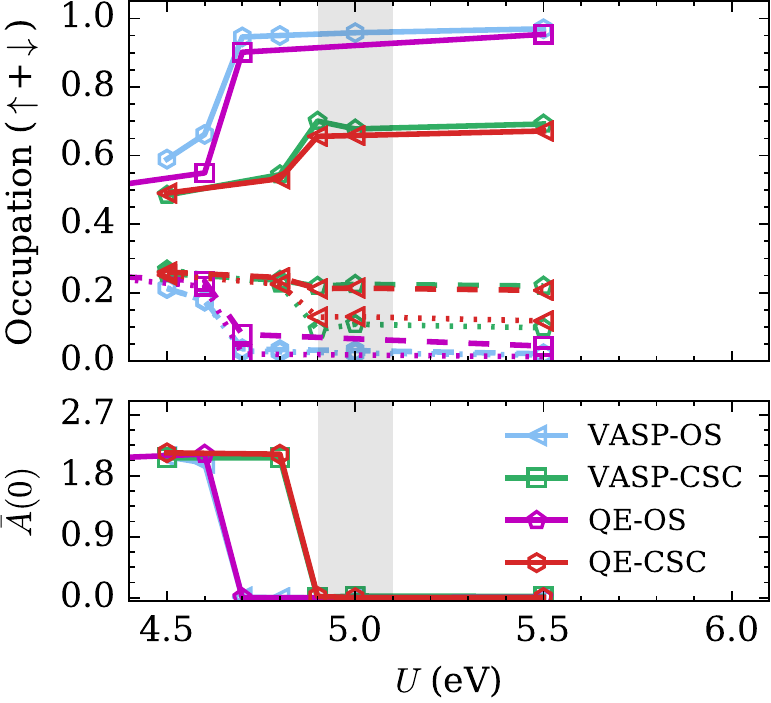}
  \caption{Adapted from Ref.~\cite{Hampel/Beck/Ederer:2020}.
  Evolution of orbital occupation (top) and spectral weight at the Fermi level $\bar{A}(0)$ (bottom) as function of interaction strength $U$.
  Comparison between \ac{OS} and \ac{CSC} calculations between \ac{VASP} and the \ac{QE}+\ac{W90} route.
  New data of the \ac{CSC} implementation is marked as red lines and symbols.
  Solid, dashed and dotted lines refer to the three \ttg{} orbitals (top panel).
  }
  \label{fig:example_cvo}
\end{figure}
\begin{figure}
  \centering
  \includegraphics[width=\linewidth]{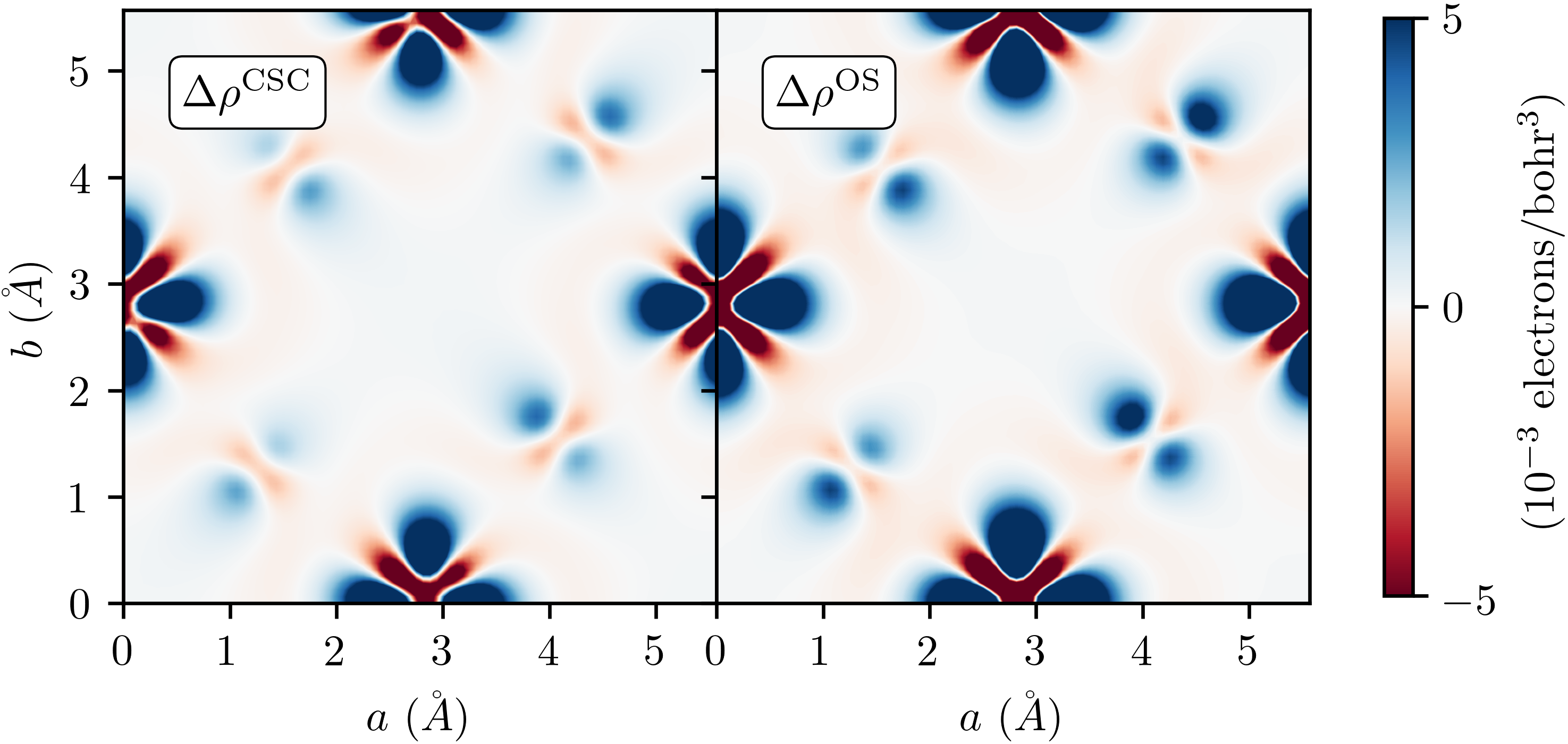}
  \caption{
  Isosurface plot of the charge redistribution in the VO$_2$ layer (averaged over contributions $\pm 3.75\%$ of the $z-$axis lattice constant) in \cvo{} in $Pbnm$ symmetry at  $U = \SI{5.0}{\textrm{eV}}$.
  Comparison between $\Delta\rho^{\mathrm{CSC}}$ (left) and $\Delta\rho^{\mathrm{OS}}$ (right) as defined in eq.~\ref{eq:deltarho_csc} and eq. \ref{eq:deltarho_os}, respectively.
  }
  \label{fig:cdens_cvo}
\end{figure}
\begin{figure*}
  \begin{tikzpicture}
    \node[anchor=south west, inner sep=0] (image) at (0, 0) {
    \includegraphics[width=.33\linewidth]{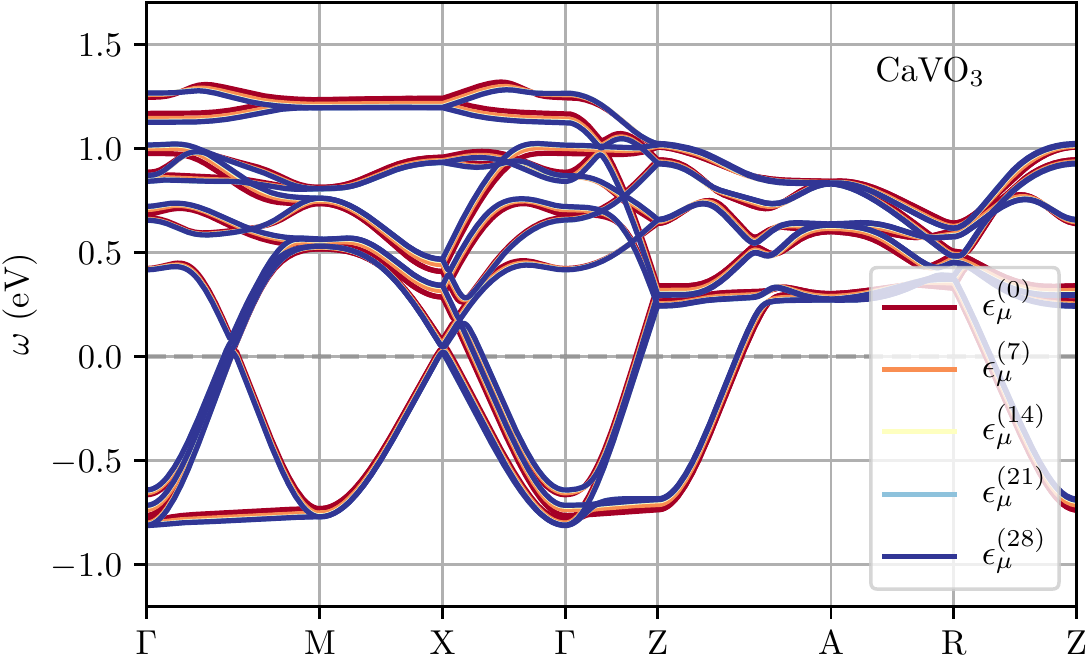}\hskip+0.005\linewidth
    \includegraphics[width=.33\linewidth]{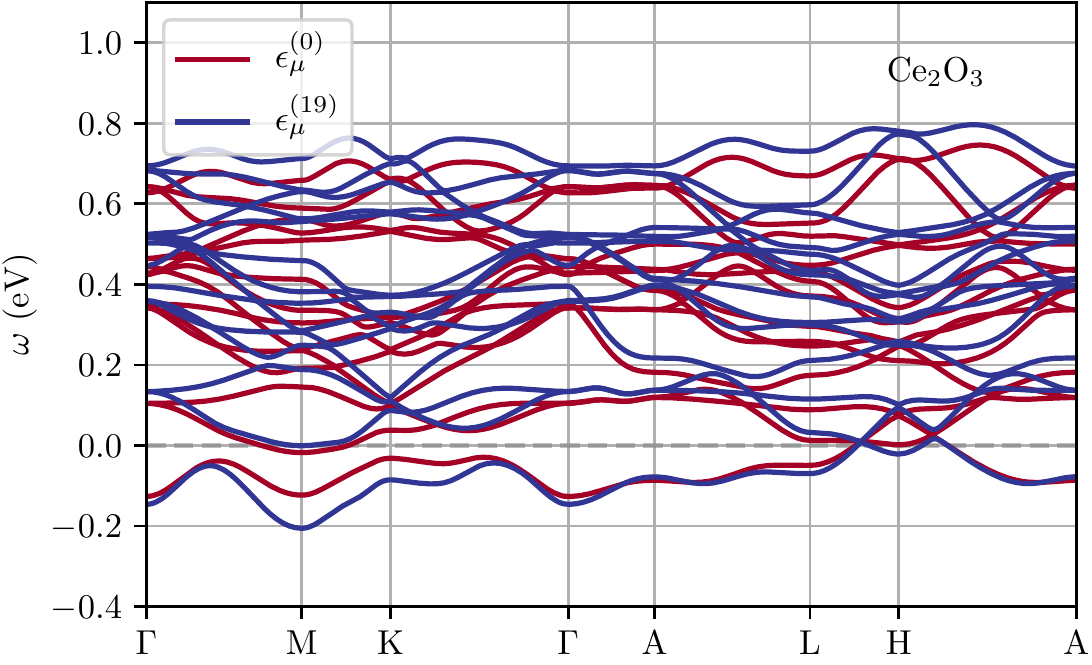}\hskip+0.005\linewidth
    \includegraphics[width=.33\linewidth]{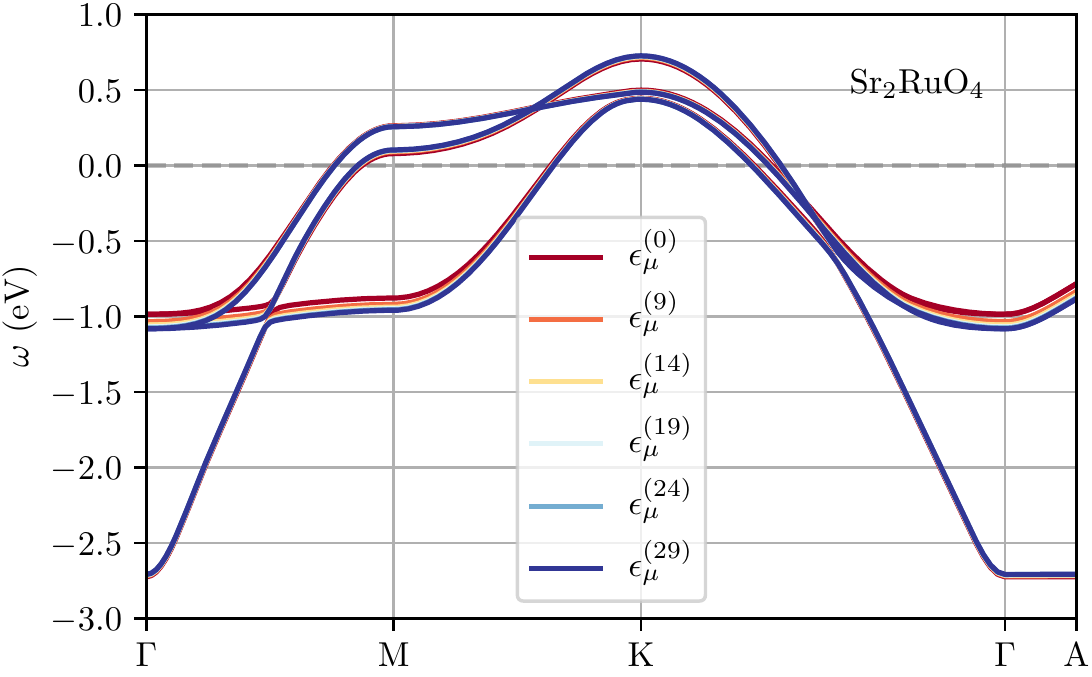}
    };
    \begin{scope}[x={(image.south east)}, y={(image.north west)}]
      \node at (0.015,1.){a)};
      \node at (0.345,1.){b)};
      \node at (0.675,1.){c)};
    \end{scope}
  \end{tikzpicture}
  \caption{
  Bandstructure of \ac{TB} Hamiltonian in band basis $\hat\epsilon^{\mathrm{TB}}_{\mu}(\ke)$ (eq.~\ref{eq:TBHamiltonian}) as function of \ac{CSC} iteration for a) \cvo{}, b) \ceo{} (experimental lattice constant) and c) \sro{}.
  Note that only $\hat\epsilon^{(0)}_{\mu}(\ke)$ represents the (Wannierized) bandstructure derived from a physical charge density, i.e. the ground state charge density $\rho^0(\ve)$.
  }
  \label{fig:bandstructure}
\end{figure*}
We first investigate the strain-induced increase of orbital polarization in the orthorhombic \done{} correlated metal \cvo{}.
Comparing \acf{OS} (i.e. non charge self-consistent) versus \ac{CSC} calculations, the authors in Ref.~\cite{Hampel/Beck/Ederer:2020} reported a shift of the critical \umit{} for the \ac{MIT} of \SI{0.2}{eV} upon introducing charge self-consistency, rendering the \ac{CSC} solution less correlated, see blue and green lines and markers in Fig.~\ref{fig:example_cvo}.
While in Ref.~\cite{Hampel/Beck/Ederer:2020} the \ac{CSC} calculations were performed only with the \ac{VASP}~\cite{Kresse:1993bz, Kresse:1996kl, Kresse:1999dk}, a comparison for \ac{OS} to \ac{QE} + \ac{W90} (in the Wannier-mode) was presented (purple lines and markers).
Here, we extend this study to complete the comparison by including the \ac{CSC} solution using \ac{QE} + \ac{W90}, see red lines and markers in Fig.~\ref{fig:example_cvo}. 
As in the previous study, all calculations were performed on a minimal \ttg{} model at $\beta = \SI{40}{\textrm{(eV)}^{-1}}$, using the standard Hubbard-Kanamori interaction~\cite{Vaugier/Jiang/Biermann:2012} with varying $U$ and fixed $J = \SI{0.65}{\textrm{eV}}$, and the continuous-time QMC hybridization-expansion solver~\cite{Gull:2011} implemented in \textsc{TRIQS/cthyb}~\cite{Seth2016274}.
Note that this model is an example of an isolated subspace as defined in case 2 of Sec.~\ref{subsubsec:proj}.
As demonstrated in Ref.~\cite{Hampel/Beck/Ederer:2020}, the comparison between the \ac{VASP}-internal projection to localized orbitals and the down-folding via \ac{W90} (\ac{DMFT} in ``Wannier mode") shows excellent agreement for the \ac{OS} case.
With our implementation we confirm that a reduction of orbital polarization, in this case occupation of $d_{\mathrm{xy}}$ with respect to $d_{\mathrm{xz}}$/$d_{\mathrm{yz}}$, occurs in \ac{CSC} calculations with a shift of \SI{0.2}{eV}, i.e in excellent agreement to previous results.

In Fig.~\ref{fig:cdens_cvo} we compare the changes in the charge density with respect to the \ac{KS} ground state density $\rho^{\mathrm{KS}}_0(\ve)$ both for \ac{CSC} and \ac{OS} calculations, i.e. $\Delta\rho^{\mathrm{CSC}}(\ve)$ and $\Delta\rho^{\mathrm{OS}}(\ve)$, respectively, defined as
\begin{align}
    \Delta\rho^{\mathrm{CSC}}(\ve) &= \rho^{\mathrm{CSC}}(\ve) - \rho^{\mathrm{KS}}_0(\ve) \label{eq:deltarho_csc}\,\\
    \Delta\rho^{\mathrm{OS}}(\ve) &= \rho^{\mathrm{OS}}(\ve) - \rho^{\mathrm{KS}}_0(\ve)\label{eq:deltarho_os}\,.
\end{align}
Both panels correctly represent the increase in orbital polarization, with a positive value for the $d_{\mathrm{xy}}$ and a reduction for $d_{\mathrm{xz}}$/$d_{\mathrm{yz}}$ orbitals, as compared to the ground state density.
The comparison between \ac{CSC} and \ac{OS} in the left and right panel, respectively, further confirms that the magnitude of orbital polarization is overestimated in \ac{OS} calculations (compare Fig.~\ref{fig:example_cvo}).
Note that $\rho^{\mathrm{OS}}(\ve)$ was computed using eq.~\ref{eq:hatdeltaN} based on a fully converged \ac{OS} calculation.
We show the changes in the Wannier Hamiltonian $\hat\epsilon_{\mu}$ (eq.~\ref{eq:TBHamiltonian}) in Fig.~\ref{fig:bandstructure}a).
The visibly small modifications demonstrate that no major reconstruction takes place, that would require attention in the construction of the \ac{TB} Hamiltonian.

\paragraph{Total energy calculation of \ceo{}}
\label{para:example_ceo}

\begin{figure}
  \centering
  \includegraphics[width=\linewidth]{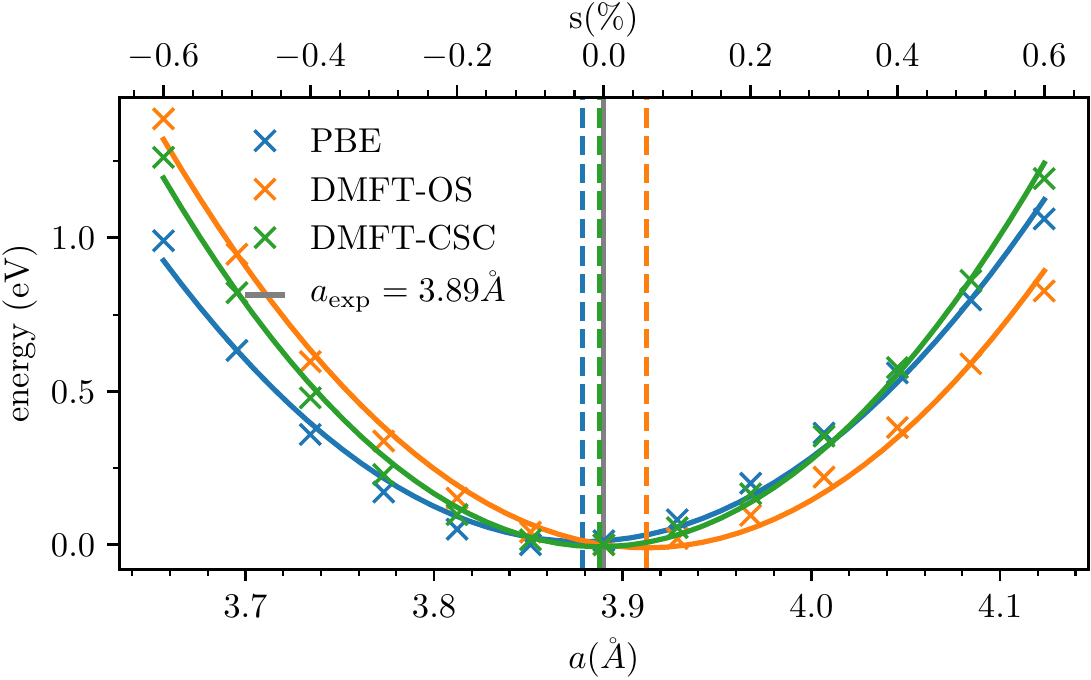}%
  \caption{Total energy calculations for hydrostatic strain in \ceo{} using the PBE functional in \ac{DFT} (blue), and \ac{OS}- and \ac{CSC}-type \ac{DMFT} calculations in orange and blue, respectively.
  The experimental lattice parameter is shown as vertical grey line. 
  }
  \label{fig:example_ceo}
\end{figure}
\begin{figure}
  \centering
  \includegraphics[width=.9\linewidth]{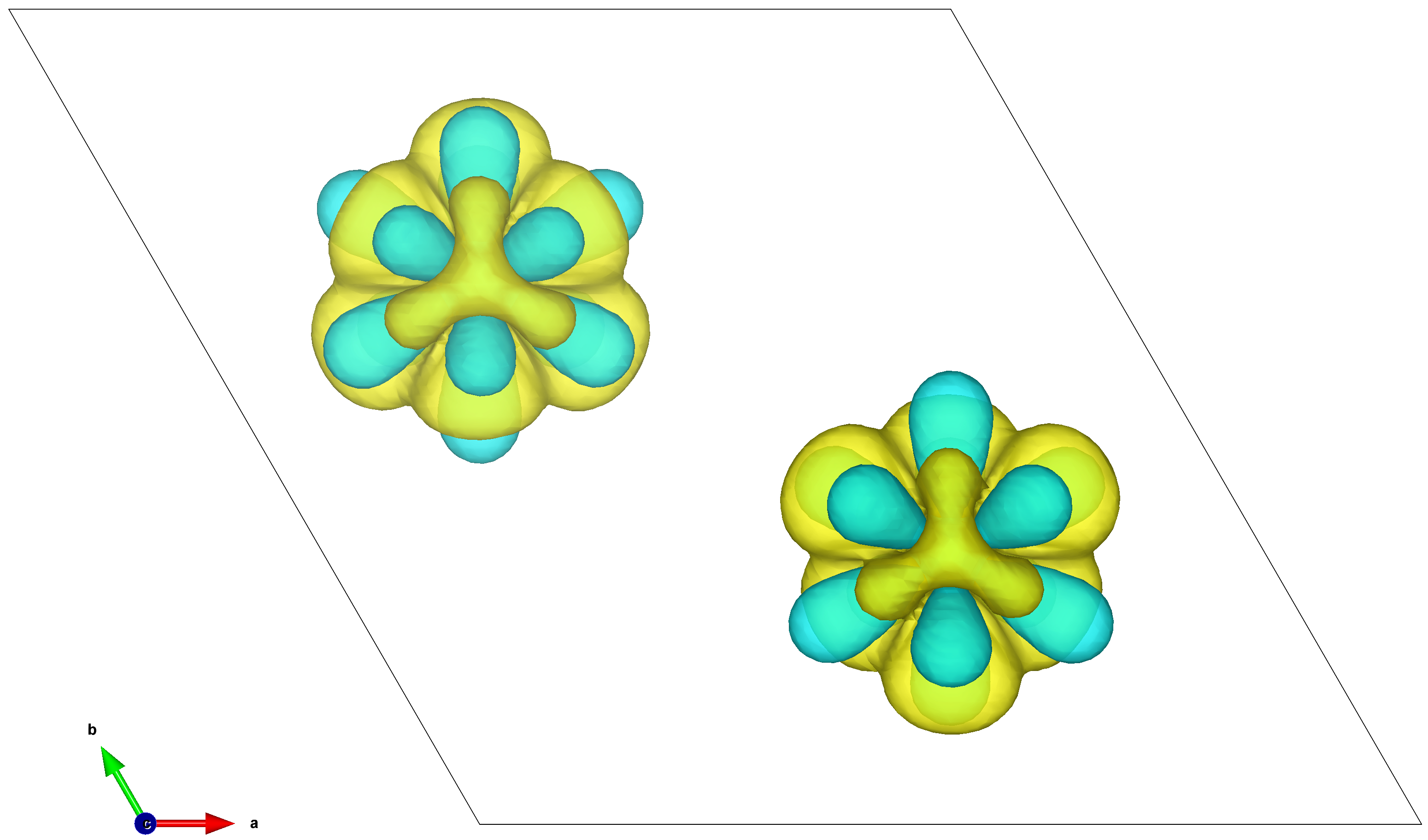}
  \caption{
  Top view of the isosurface plot of the charge redistribution $\Delta\rho^{\mathrm{CSC}}$ (eq.~\ref{eq:deltarho_csc}) in \ceo{} at an isovalue of $\SI{7.8 e-3}{electrons/bohr^{3}}$.
  Yellow and cyan represent positive and negative changes, respectively.
  }
  \label{fig:cdens_ce2o3}
\end{figure}
The second example is given by total energy calculations of the equation of state for \ceo{} as function of volume.
Here, we compute the total energy for equidistantly spaced strain values between \SIrange[range-phrase ={\text{~to~}}]{-0.6}{0.6}{\%} for a low-energy Hamiltonian based on the correlated $f$-shell.
The low-energy model constitutes a fully isolated set of bands as described in Sec.~\ref{subsubsec:proj}, case 2.
The corresponding impurity problem is solved within \ac{DMFT} with the Hubbard-I impurity solver at $\beta = \SI{2000}{\textrm{(eV)}^{-1}}$ using $U = \SI{6.46}{\textrm{eV}}$ and $J = \SI{0.46}{\textrm{eV}}$.
We use the fully localized limit as the \ac{DC} correction formula as was previously done in Ref.~\cite{Pourovskii_et_al:2007}.
As is shown in Fig.~\ref{fig:example_ceo}, the \ac{DFT} solution (blue lines and markers) slightly underestimates the equilibrium lattice constant obtained in the experiment, although it is already in very good agreement with a deviation of less than \SI{-0.5}{\%}.
The equilibrium lattice constant then shifts to \SI{\approx 0.6}{\%} in \ac{OS} \ac{DMFT} calculations (orange lines and markers).
Finally, introducing full charge self-consistency the agreement to the experiment is almost perfect (green lines and markers).
The minima of the quadratic fits are indicated by vertical dashed lines.
The changes in the non-interacting bandstructure and charge density are displayed in Fig.~\ref{fig:bandstructure}b) and Fig.~\ref{fig:cdens_ce2o3}, respectively, for the experimental volume.
In comparison to the other two materials the magnitude of the changes in the largest in \ceo{}, with a change in $\Delta\rho^{\mathrm{CSC}}$ (eq.~\ref{eq:deltarho_csc}) twice as large as for \cvo{} in Fig.~\ref{fig:cdens_cvo}.

We note that in comparison to Ref.~\cite{Pourovskii_et_al:2007} already the \ac{DFT} equilibrium is much closer to the experimental value.
This is attributed to the difference of exchange functionals from LDA to PBE~\cite{Perdew:1996iq}, which are known to underestimate/overestimate lattice parameters, respectively.
The shift to larger lattice parameters in the \ac{OS} \ac{DMFT} solution can be understood in terms of the expansion of $4f$ orbitals due to the local interaction, similar as in \ac{DFT}$+U$.
Finally, the \ac{CSC} solution lies in between the \ac{DFT} and the \ac{OS} results, which one might intuitively expect if one considers the self-consistent solution as a mixture of the two extrema, the \ac{DFT} solution on the one hand and the \ac{OS} \ac{DMFT} solution on the other hand.

\paragraph{Quasiparticle mass renormalization in \sro{}}
\label{para:example_sro}

\begin{figure}
  \centering
  \includegraphics[width=\linewidth]{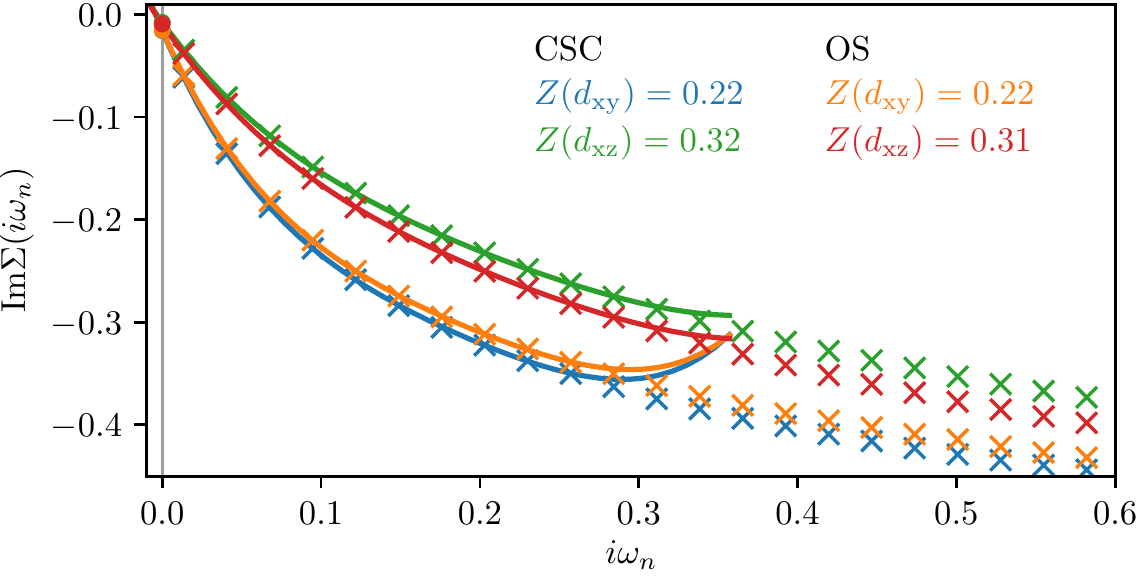}%
  \caption{Comparison between \ac{OS} (orange and red) and \ac{CSC} (blue and green) of the imaginary part of the self-energy for the $d_{\mathrm{xy}}$ and $d_{\mathrm{xz}}$ orbitals.
  Crosses correspond to numerical results, solid lines are fits of the first ten Matsubara frequencies to a polynomial of fourth order.
  }
  \label{fig:example_sro}
\end{figure}

We finally turn to \sro{}, a metal displaying clear signatures of strong electronic correlations. 
Previous DMFT studies have established that the Hund's rule coupling (intra-atomic exchange) is responsible for these electronic correlations~\cite{mravlje_2011,Georges/Medici/Mravlje:2013}, the proximity of a van Hove singularity also playing an important role~\cite{kugler_2020}.
Most previous studies have however been performed in OS mode without ensuring full charge self-consistency, with only a couple of published works~\cite{mravlje_2016,deng_2016} mentioning that, indeed, charge self-consistency does not lead to significant modifications - a fact which is perhaps to be expected given the highly itinerant character and rather large bandwidth of this metal. 
Here, we document this in more details. 
Due to overlap of the Ru $d_{\mathrm{xy}}$ orbital with oxygen $p$ states, the construction of a Wannier model for the Ru \ttg{} orbitals is an example of an entangled subspace as defined in case 3 of Sec.~\ref{subsubsec:proj}.
We focus here on key information about the nature of quasiparticles in this system, which can be extracted from a DMFT calculation, namely the quasiparticle mass renormalization $Z^{-1} = m^*/m \approx 1 - \frac{Im\Sigma(\im \omega_n)}{\im \omega_n}|_{\omega_0}$. 
The impurity problem with Hubbard-Kanamori interactions is solved with the continuous-time QMC hybridization-expansion solver with the interaction parameters set as $U = \SI{2.3}{eV}$ and $J = \SI{0.4}{eV}$.
In the \ac{OS} calculation at $\beta = \SI{232}{\textrm{(eV)}^{-1}}$ ($T= 50$~K) the four electrons in the $d$-shell are equally distributed with very low polarization of $\approx 1.29/1.35$ between the $d_{\mathrm{xy}}$ and $d_{\mathrm{xz}}$/$d_{\mathrm{yz}}$ orbitals, respectively.
The corresponding self-energies are shown in Fig.~\ref{fig:example_sro} (orange and red, respectively).
Fitting the first ten Matsubara frequencies to a polynomial of fourth order, the extracted slope at zero frequency results in an inverse quasiparticle mass renormalization of $Z=0.22/0.31$ for the two orbitals, respectively.
As expected, performing \ac{CSC} calculations does not lead to a drastic change of these results.
The orbital polarization changes less than \SI{2}{\%}.
Similarly, the inverse quasiparticle mass renormalization (blue and green in Fig.~\ref{fig:example_sro}) increases minimally to $0.32$ for the $d_{\mathrm{xz}}$/$d_{\mathrm{yz}}$ orbitals.
This is to be expected from the very small changes of the non-interacting Hamiltonian $\hat\epsilon^{\mathrm{TB}}_{\mu}(\ke)$ as shown in Fig.~\ref{fig:bandstructure}c).

\section{Conclusion}
\label{sec:conclusion}

In conclusion, we have presented a new implementation for fully \acl{CSC} \ac{DFT}+\ac{DMFT} calculations using a combination of the \acl{QE} (v.7.0), \acl{W90} and \textsc{TRIQS/DFTTools} (v.3.1) software packages. 
Our approach unifies the Wannier-based and projector-based embedding methods. We note that 
in our approach the projectors are not necessarily confined to spheres around the projection center. For the sake of clarity, we have introduced a revised naming convention, Bloch- versus Wannier-mode, referring to the basis in which the \ac{DMFT} equations are formulated in as a  fundamental distinction.

While the overall implementation is tailored for the particular combination of codes, the path through \acl{W90} is more general and can be combined with other \ac{DFT} codes as well as \acl{TB} approaches.
It thereby allows to utilize standard tools included in or bridged to \acl{W90}, making this approach more flexible and appealing to a broader community.
In comparison to a simpler projection scheme in the typical ``projector''-based approaches, it provides more control of the correlated subspace basis, allowing the user to define parameters of spread minimization or finding an optimally connected subspace in case of entanglement.
A simple visual interface further allows to check the legitimacy of the subspace in representing the \acl{KS} states correctly, which is often inaccessible in ``projector''-based formalisms.

We benchmarked the implementation for three different systems and showed that it produces excellent agreement to previously reported and published results based on existing implementations of different codes.
The workflow is fully open-source and \ac{MPI}-parallelized, which makes it ideally suited for a broader community and for studying larger systems.

\section*{Acknowledgements}

S.B. is grateful to P. Giannozzi, P. Delugas and I. Timrov for helpful comments and insightful discussions on the implementation in \ac{QE}.
This work was supported by ETH Zurich and the Swiss National Science Foundation through NCCR-MARVEL.
Calculations were performed on the cluster ``Piz Daint'' hosted by the Swiss National Supercomputing Centre.
The Flatiron Institute is a division of the Simons Foundation.

\appendix

\section{Differences \ac{TB} and Bloch formulation of \ac{DMFT} equations}
\label{sec:appendixA}

Here, we demonstrate that writing the \ac{DMFT} equations in orbital basis is equivalent to a formulation in Bloch basis if and only if the projector functions $\hat P$ are unitary, in which case one can skip the upfolding of the self-energy in eq.~\ref{eq:upfolding} and trivially obtain the $\ke$-dependence of the self-energy.

In operator notation the lattice Green's function is defined as
\begin{align}
    \hat G(\ke, \im \omega_n)^{-1} = (\im \omega_n + \mu)\mathbb{\hat 1}  - \hat \epsilon(\ke) - \Delta \hat\Sigma(\ke, \im \omega_n)\,.
\end{align}
We express eq.~\ref{eq:downfolding} without $\ke{}$-summation in operator notation as
\begin{align}
    \hat G^{\text{TB}} &= \hat P\, \hat G^{\text{KS}} \hat P^{\dagger} %
    = \hat P \left[ (\hat G^{\text{KS}})^{-1} \right]^{-1} \hat P^{\dagger}\,,\nonumber\\
    &= \left[ (\hat P^{\dagger})^{-1} (\hat G^{\text{KS}})^{-1} \hat P^{-1}\right]^{-1}\,,
\end{align}
where TB and KS refer to the lattice Green's function in \ac{TB} and \ac{KS} basis, respectively.
If $\hat P$ is unitary, the last line reduces to
\begin{align}
    \hat G^{\text{TB}} &= \left[\hat P\, (\hat G^{\text{KS}})^{-1} \hat P^{\dagger}\right]^{-1} \,,
\end{align}
in which case the projector functions can be applied just once to down-fold the \ac{KS} Hamiltonian as in eq.~\ref{eq:downfolding_hloc},
\begin{align}
    \hat G^{\text{TB}}(\ke, \im \omega_n)^{-1} ={}& (\im \omega_n + \mu)\mathbb{\hat 1} - \hat P(\ke) \hat \epsilon^{\text{KS}}(\ke) \hat P^{\dagger}(\ke)\nonumber\\
    &- \hat P(\ke) \Delta \hat\Sigma(\ke, \im \omega_n) \hat P^{\dagger}(\ke)\,\nonumber\\
    ={}& (\im \omega_n + \mu)\mathbb{\hat 1}  - \hat \epsilon^{\text{TB}}(\ke) - \Delta \hat\Sigma^{\text{TB}}(\ke, \im \omega_n)\,,
\end{align}
Only in this case the impurity self-energy obtains a trivial $\ke$-dependence as
\begin{align}
    \Delta \hat\Sigma^{\text{TB}}(\ke, \im \omega_n) = \sum_{mm'} \Delta \Sigma_{mm'}(\im \omega_n) \mathbb{1}_{mm'}\,.
\end{align}
This formalism is readily extended to multiple impurity sites $i$ with block-diagonal impurity Hamiltonians.
The resulting self-energy in $\mathcal{C}$ is block-diagonal, too, meaning that the up-folding in eq.~\ref{eq:upfolding} can be done independently for each block/site, i.e. the matrix multiplication separates into a sum over blocks/sites $i$ with corresponding rows of the projector functions.

\bibliographystyle{elsarticle-num}
\bibliography{bibfile.bib}

\end{document}